\documentclass[%
aip,
amsmath,amssymb,
reprint,%
]{revtex4-1}

\usepackage{graphicx}
\usepackage{dcolumn}
\usepackage{bm}
\usepackage{amsmath, amsthm, amsxtra, amsfonts, amssymb,amscd,mathtools}
\usepackage{latexsym}
\usepackage{color}

\usepackage[utf8]{inputenc}
\usepackage[T1]{fontenc}
\usepackage{mathptmx}
\usepackage{etoolbox}
\usepackage{autobreak}
\usepackage[normalem]{ulem}

\draft 

\makeatletter
\def\@email#1#2{%
 \endgroup
 \patchcmd{\titleblock@produce}
  {\frontmatter@RRAPformat}
  {\frontmatter@RRAPformat{\produce@RRAP{*#1\href{mailto:#2}{#2}}}\frontmatter@RRAPformat}
  {}{}
}%
\makeatother

\begin{document}


\title{Periodic forces combined with feedback induce quenching in a bistable oscillator} 



\author{Yusuke Kato}
  \email[]{yuukatou-tky@umin.ac.jp}
\author{Hiroshi Kori}
\affiliation{
Department of Complexity Science and Engineering, Graduate School of Frontier Sciences, The University of Tokyo, Kashiwa, Chiba 277-8561, Japan
}

\date{\today}

\begin{abstract}
    The coexistence of an abnormal rhythm and a normal steady state is often observed in nature (e.g., epilepsy). Such a system is modeled as a bistable oscillator that possesses both a limit cycle and a fixed point. 
    Although bistable oscillators under several perturbations have been addressed in the literature, the mechanism of oscillation quenching, a transition from a limit cycle to a fixed point, has not been fully understood.
    In this study, 
    we analyze quenching using the extended Stuart-Landau oscillator driven by periodic forces. Numerical simulations suggest that the entrainment to the periodic force induces the amplitude change of a limit cycle. By reducing the system with an averaging method, we investigate the bifurcation structures of the periodically-driven oscillator. 
    We find that oscillation quenching occurs by the homoclinic bifurcation when we use a periodic force combined with quadratic feedback. In conclusion, we develop a state-transition method in a bistable oscillator using periodic forces, which would have the potential for practical applications in controlling and annihilating abnormal oscillations. 
    Moreover, we clarify the rich and diverse bifurcation structures behind periodically-driven bistable oscillators, which we believe would contribute to further understanding the complex behaviors in non-autonomous systems. 
\end{abstract}


\maketitle 

\begin{quotation}
    From cell cycles to the rotation of the earth, rhythmic behaviors are commonly observed in nature. Some oscillations are associated with abnormal states and often coexist with other normal equilibrium states. Motivated by this, we focus on a bistable oscillator and investigate whether oscillation quenching is induced by periodic forces. By focusing on the synchronization state between the oscillator and the periodic force, we apply an averaging approximation and perform a bifurcation analysis. We find that a periodic force combined with quadratic feedback induces oscillation quenching. Our findings present a clue for the practical application in annihilating undesirable oscillations, as well as provide insight into the understanding of diverse dynamics observed in non-autonomous systems. 
\end{quotation}

\section{Introduction}
Among various oscillatory phenomena in nature, some oscillations are associated with abnormal and undesirable states. For example, in patients with Parkinson's disease, oscillatory electrical activities are recorded in specific brain areas \cite{brown2001dopamine}. These oscillatory activities reflect pathological brain states in Parkinson's disease \cite{brown2003oscillatory,little2014functional} and are suppressed by medical \cite{brown2001dopamine} or electrical treatments \cite{eusebio2011deep}. Another example of abnormal oscillation is observed in epileptic seizures. In a patient with absence seizure, a specific type of epileptic seizures, a 3 Hz oscillation with a large amplitude is typically recorded with electroencephalogram \cite{kessler2019practical}. 

Abnormal oscillations often co-exist with normal steady states, and thus they are modeled as a bistable oscillator, i.e., an oscillator that has both a stable limit cycle and a stable fixed point. For example, in the modeling of epileptic seizures, various bistable oscillators are used to describe both the seizure (a stable limit cycle) and the normal (a stable fixed point) brain states \cite{jirsa2014nature,kalitzin2010stimulation,suffczynski2004dynamics,suffczynski2005epileptic,benjamin2012phenomenological}. The annihilation of abnormal rhythms by adding external perturbations and inducing oscillation quenching, a state transition from a limit cycle to a fixed point, is expected to contribute to the treatment of diseases \cite{chang2020falling}. 

Previous studies have addressed state transitions in bistable oscillators. The perturbations used in these studies include noise \cite{zakharova2010stochastic,xu2011stochastic,suffczynski2004dynamics,kalitzin2010stimulation}, feedback \cite{kalitzin2010stimulation}, pulse \cite{kalitzin2010stimulation}, and the combination of these \cite{Sun2015}. Recently, Chang et al. applied a stochastic optimization algorithm to find an optimal shape of external stimuli that induce oscillation quenching \cite{chang2020falling,chang2021falling}. However, several open questions remain in these previous studies. First, transitions from a fixed point to a limit cycle, the inverse transition of our interest, were observed in some of these studies \cite{zakharova2010stochastic,xu2011stochastic,Sun2015}. In addition, even though quenching was achieved in other studies, they did not analyze the mechanism of quenching theoretically \cite{kalitzin2010stimulation,chang2020falling,chang2021falling}. The lack of analytical studies that treat quenching prevents the systematic understanding of state-transition methods in a bistable oscillator. 

In this article, we analyze a bistable oscillator driven by periodic forces. By adding periodic forces, the system becomes non-autonomous, which would enrich the dynamics but also make the analysis difficult.
Periodically-driven oscillators have been analyzed in several previous studies in relationship to fluid dynamics \cite{le2001hysteresis,thompson2004stuart,fang2004some}, multi-body systems \cite{kedia2023drive}, or pure theoretical interests \cite{mettin1993bifurcation,yonkeu2017effects,wawrzynski2021duffing,koch2024ghost}. However, most of these studies \cite{le2001hysteresis,thompson2004stuart,fang2004some, mettin1993bifurcation,wawrzynski2021duffing,koch2024ghost} addressed a monostable oscillator, in which either a fixed point or a limit cycle is stable without periodic forces. Moreover, although frequency resonance curves are obtained near the supercritical Hopf bifurcation \cite{le2001hysteresis,thompson2004stuart}, few studies focus on the amplitude response of a stable limit cycle with a large amplitude. 
Aiming to clarify whether periodic forces can induce oscillation quenching, we analyze the amplitude response of a bistable oscillator driven by several types of periodic forces. 

The remainder of this paper is organized as follows: In Sec. \ref{sec_esl}, we introduce an extended Stuart-Landau oscillator [Eq. \eqref{esl}] as a bistable oscillator. In Sec. \ref{sec_esl_p}, the amplitude response under an additive periodic force is described. 
Numerical simulations suggest that 1:1 synchronization between the frequencies of the oscillator and the periodic force occurs. By using an averaging method \cite{pikovsky2002synchronization,Guckenheimer1983}, we reduce the system into an autonomous one and perform a bifurcation analysis. In Sec. \ref{sec_esl_mp1} and \ref{sec_esl_mp2}, we apply the same analyses to the oscillators under multiplicative periodic forces [Eqs. \eqref{esl_mp} and \eqref{esl_mp2}]. In particular, we find that oscillation quenching occurs when the product of periodic force and quadratic feedback is used as a perturbation. Finally, in Sec. \ref{sec_dis}, we describe the discussion and conclusion.  

\section{Model: Extended Stuart-Landau oscillator}
\label{sec_esl}
\begin{figure}
    \centering
    \includegraphics[width = 1.\linewidth]{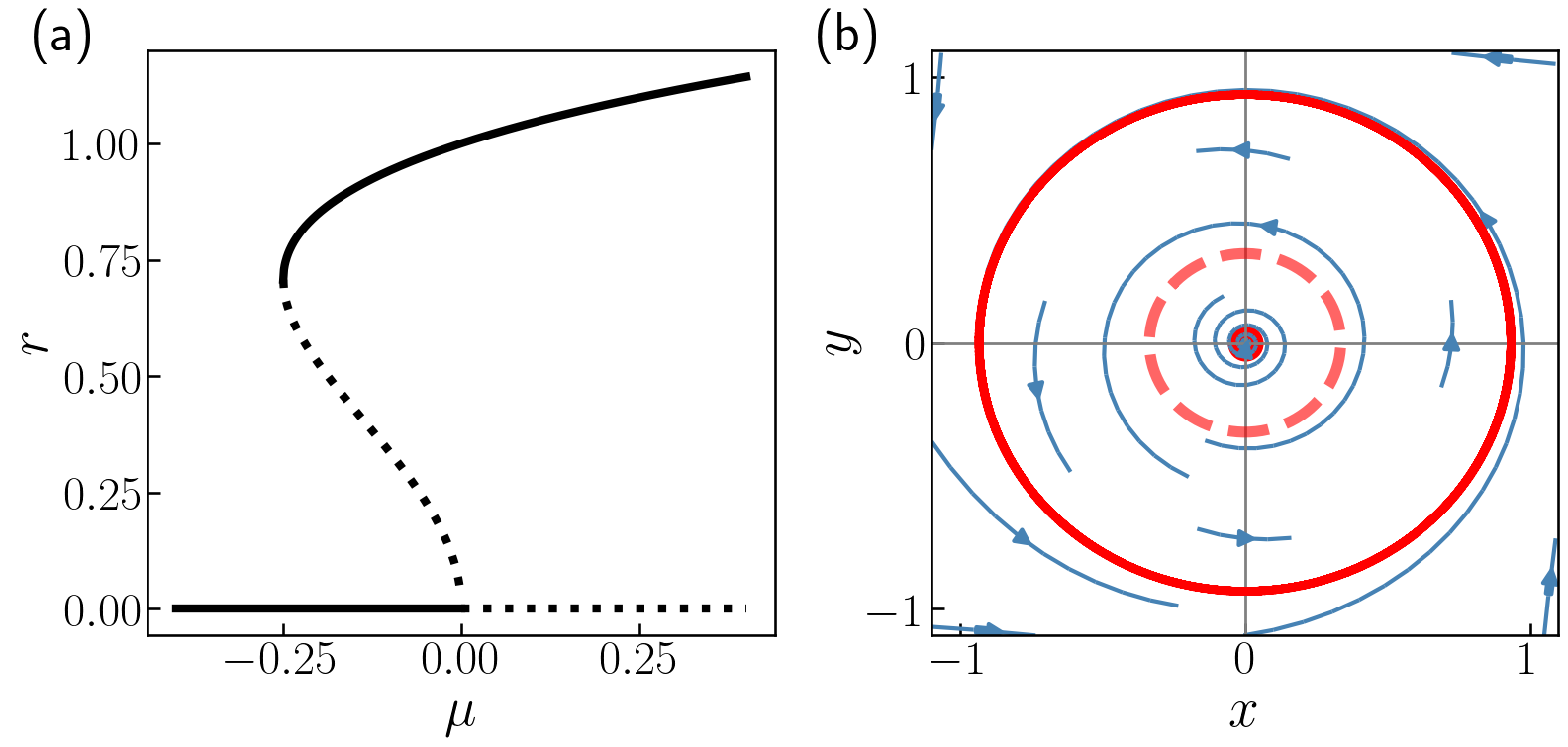}
    \caption{(a) A bifurcation diagram of Eq. \eqref{esl_r}. The solid and dotted lines represent the stable and unstable branches, respectively. (b) A typical phase plane of Eq. \eqref{esl} when $a=b=1.0,\: \mu=-0.1$. The red dot shows a stable fixed point at the origin. The solid and dashed red lines show the stable and unstable limit cycles, respectively.}
    \label{esl_bif_pp}
\end{figure}

We consider the following extended Stuart-Landau oscillator:
\begin{subequations}
    \label{esl}
    \begin{align}
        \dot x &= \mu x - a y + (x^2 + y^2)(x - b y) - (x^2 + y^2)^2 x, \label{slx}\\
        \dot y &= a x + \mu y + (x^2 + y^2)(b x + y) - (x^2 + y^2)^2 y, \label{sly}
    \end{align}
\end{subequations}
which can be rewritten as 
\begin{equation}
    \label{esl_comp}
    \dot z = (\mu + ia)z + (1 + ib)|z|^2 z - |z|^4 z,
\end{equation}
where $z \coloneqq x + iy$. 
A polar expression of Eq. \eqref{esl_comp} [i.e., $z \eqqcolon r(t) e^{i \theta(t)}$] yields
\begin{subequations}
    \begin{align}
        \dot r &= \mu r + r^3 - r^5, \label{esl_r}\\
        \dot \theta &= a + b r^2, \label{esl_theta}
    \end{align}
\end{subequations}
which implies that system \eqref{esl} has both a stable limit cycle and a stable fixed point if and only if $-\frac{1}{4} < \mu < 0$. Figure \ref{esl_bif_pp} shows a bifurcation diagram of Eq. \eqref{esl_r} and a typical phase plane of Eq. \eqref{esl}. 
Note that the amplitudes and frequencies of stable and unstable limit cycles are, respectively, 
\begin{equation}
    r_{\rm s} = \frac{\sqrt{2 + 2 \sqrt{1 + 4 \mu}}}{2}, \quad \omega_{\rm s} = a + \frac{b}{2}(1+\sqrt{1+4\mu})
\end{equation}
and
\begin{equation} 
    r_{\rm u}  = \frac{\sqrt{2 - 2 \sqrt{1 + 4 \mu}}}{2}, \quad \omega_{\rm u} = a + \frac{b}{2}(1 - \sqrt{1+4\mu}).
\end{equation} 

In what follows, we fix $\mu = -0.1$ so that the system \eqref{esl} can be viewed as a bistable oscillator. We also fix $a = b = 1.0$ for all the numerical simulations in this article. 

\section{Additive periodic force}
\label{sec_esl_p}
Using model \eqref{esl}, we analyze the amplitude response of the stable limit cycle under three types of (nearly) periodic perturbations. In this section, we apply an additive periodic force to only the first equation \eqref{slx}. The system is 
\begin{subequations}
    \label{esl_p}
    \begin{align}
        \dot x &= \mu x - a y + (x^2 + y^2)(x - b y) - (x^2 + y^2)^2 x + A \cos \Omega t, \label{slpx}\\
        \dot y &= a x + \mu y + (x^2 + y^2)(b x + y) - (x^2 + y^2)^2 y, \label{slpy}
    \end{align}
\end{subequations}
which is equivalent to 
\begin{equation}
    \label{esl_comp_p}
    \dot z = (\mu + ia)z + (1 + ib)|z|^2 z - |z|^4 z + A \cos \Omega t.
\end{equation}

\subsection{Numerical simulation}
\begin{figure}[ht]
    \centering
    \includegraphics[width = 1.\linewidth]{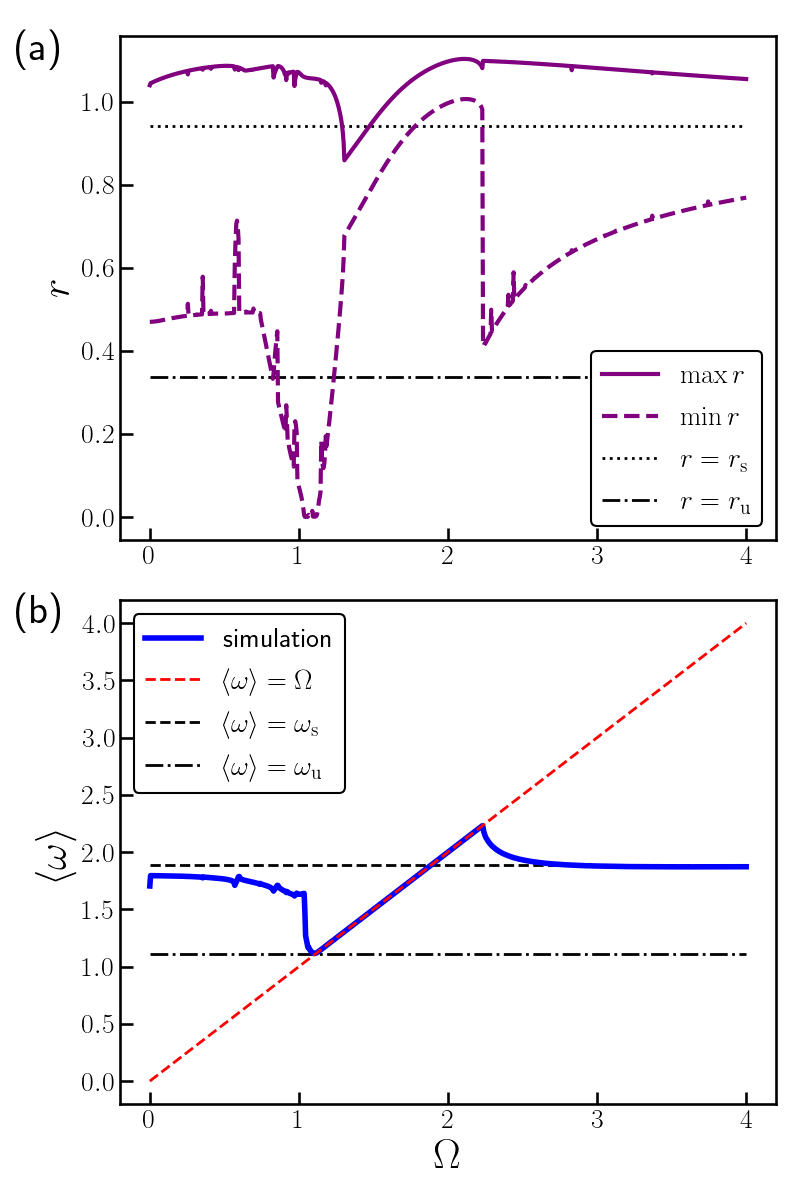}
    \caption{The amplitude (a) and frequency (b) response curves of Eq. \eqref{esl_p}. Panel (a) shows the maxima and minima of the amplitude $r$, whereas panel (b) plots the long-time average of the frequency of the oscillator $\langle \omega \rangle$. In both panels, the horizontal axis represents $\Omega$ (the frequency of periodic force). Note that panel (b) suggests that the entrainment to the periodic force occurs when $\Omega \in [\Omega_{\mathrm l}, \Omega_{\mathrm u}]$ with $\Omega_{\mathrm l} \simeq 1.12$ and $\Omega_{\mathrm u} \simeq 2.23$. For numerical simulation, we set $A = 0.5, \: x(0)=3.0,\: y(0)=0.0$ and $t_{\rm stim} = 500$, where $t_{\rm stim}$ denotes the time when the periodic force is added. 
    }
    \label{reso_esl_freq}
\end{figure}

The amplitude and frequency response curves of Eq. \eqref{esl_p} are shown in Figs. \ref{reso_esl_freq} (a) and (b), respectively. 

Figure \ref{reso_esl_freq} (a) plots the maxima and minima of the amplitude $r$, which is given by 
\begin{equation}
    \label{def_r}
    r(t) \coloneqq |x(t) + i y(t)|,
\end{equation}
against different values of $\Omega$. By noting that $\max r = \min r = r_\mathrm{s}$ in the stable limit cycle of the unperturbed system \eqref{esl}, we see that the periodic force changes the shape of trajectories, breaking the rotational symmetry of the original system \eqref{esl}. In particular, we find that $\min r \simeq 0$ for some values of $\Omega$, indicating that the trajectory passes near the origin. 

In Fig. \ref{reso_esl_freq} (b), we plot $\langle \omega \rangle$, the long-time average of the frequency of the oscillator. Here, $\omega$ denotes the instantaneous angular frequency, which is given by
\begin{equation}
    \omega(t) \coloneqq \frac{d}{dt} \arg [x(t)+iy(t)], 
\end{equation}
and $\langle \cdot \rangle$ represents a long-time average. We find that 1:1 synchronization between the oscillator and the periodic force (i.e., $\langle \omega(t) \rangle = \Omega$) is observed.  


\subsection{Analysis}
To analyze the amplitude response curve [Fig. \ref{reso_esl_freq} (a)], we focus on the 1:1 synchronized state observed in Fig. \ref{reso_esl_freq} (b) and perform an averaging approximation. 
\subsubsection{Averaging}
We introduce new variables $X, Y$, and $Z$ given by
\begin{equation}
    \label{comp_cv}
    Z = z e^{- i \Omega t}.
\end{equation}
Then, Eq. \eqref{esl_comp_p} is rewritten as
\begin{align}
    \dot{Z} 
    &= [\mu + i(a-\Omega)] Z + (1 + ib)|Z|^2 Z - |Z|^4 Z + A e^{- i \Omega t} \cos \Omega t \notag \\
    &= [\mu + i(a-\Omega)] Z + (1 + ib)|Z|^2 Z - |Z|^4 Z \notag \\
    & \qquad \qquad + A (\cos^2 \Omega t - i \sin \Omega t \cos \Omega t) \notag \\
    &= [\mu + i(a-\Omega)] Z + (1 + ib)|Z|^2 Z - |Z|^4 Z \notag \\
    & \qquad \qquad + \frac{A}{2}(1 + \cos 2 \Omega t - i \sin 2\Omega t). \label{esl_p_comp_cv}
\end{align}
Note that 
a fixed point in Eq. \eqref{esl_p_comp_cv} corresponds to a rotationally symmetric limit cycle with a frequency $\Omega$ in Eq. \eqref{esl_comp_p}. 

From Eq. \eqref{esl_p_comp_cv}, we assume that the time series of $Z$ has a fast oscillation component of frequency $2\Omega$. To focus on the slower dynamics of $Z$, we average Eq. \eqref{esl_p_comp_cv} over a period of the fast oscillation (i.e., $t \in [0, \pi/\Omega]$) assuming that $Z$ can be approximately considered as a constant during this period. Then, we obtain
\begin{equation}
    \dot{Z} = [\mu + i(a-\Omega)] Z + (1 + ib)|Z|^2 Z - |Z|^4 Z + \frac{A}{2}, \label{esl_p_comp_av}
\end{equation}
which is equivalent to 
\begin{subequations}
    \label{esl_p_XY}
    \begin{align}
        \dot X
        &= \mu X - (a-\Omega)Y + (X^2+Y^2)(X-bY) - (X^2+Y^2)^2 X + \frac{A}{2}, \label{esl_p_X}\\
        \dot Y
        &= (a-\Omega)X + \mu Y + (X^2+Y^2)(bX+Y) - (X^2+Y^2)^2 Y, \label{esl_p_Y}
    \end{align}
\end{subequations}
where $X \coloneqq \mathrm{Re} \: Z$ and $Y \coloneqq \mathrm{Im} \: Z$. We express the amplitude of the trajectory in the averaged system \eqref{esl_p_XY} as $\bar{r}$. Namely, we set
\begin{equation}
    \bar{r}(t) \coloneqq |X(t)+iY(t)|. 
\end{equation}

\begin{figure}
    \centering
    \includegraphics[width = 1.\linewidth]{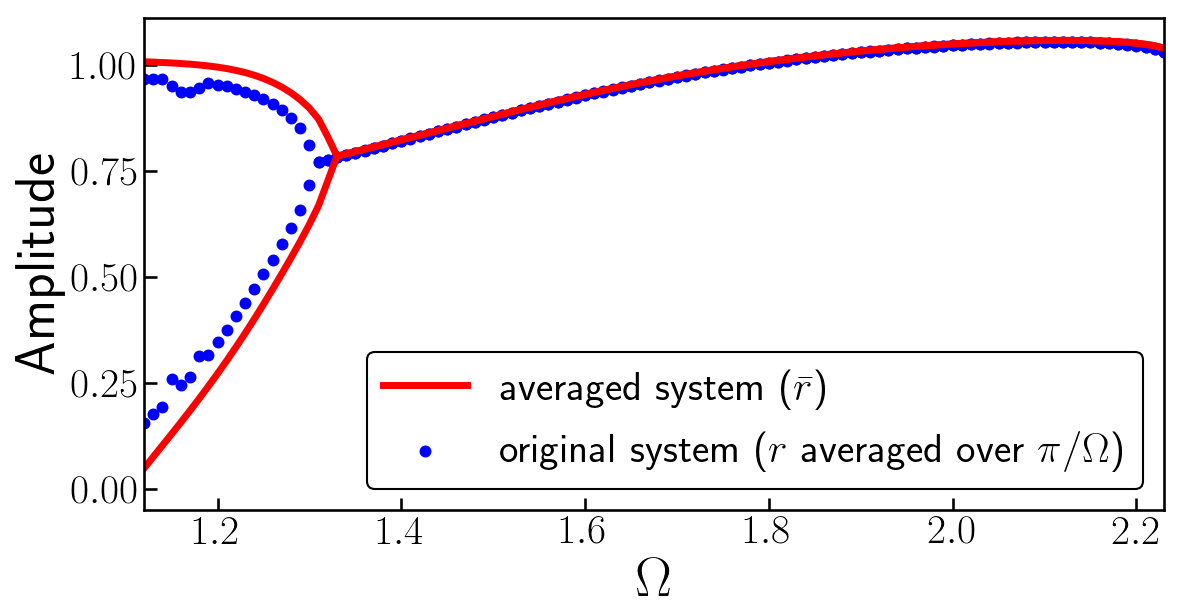}
    \caption{The comparison between the original model \eqref{esl_p} and averaged equation \eqref{esl_p_XY} when $A=0.5$. The blue dots represent the maxima and minima of $r$ averaged over $\frac{\pi}{\Omega}$, obtained by the numerical simulation of the original system \eqref{esl_p}. The red lines show the maxima and minima of $\bar{r}$, obtained by the numerical simulation of the averaged system \eqref{esl_p_XY}. }
    \label{reso_esl_freq_avr}
\end{figure}

To confirm the validity of the above averaging method, we compare the dynamics of Eq. \eqref{esl_p_XY} with those of the original model \eqref{esl_p} (Fig. \ref{reso_esl_freq_avr}). The blue dots in Fig. \ref{reso_esl_freq_avr} show the maxima and minima of $r$ averaged over $\frac{\pi}{\Omega}$, obtained from the numerical simulation of the original system \eqref{esl_p}. The red lines represent the maxima and minima of $\bar{r}$, derived from the numerical simulation of the averaged system \eqref{esl_p_XY}. They show a reasonable agreement, indicating the validity of our approximation. Thus, in the following analysis, we investigate the dynamics of the averaged system \eqref{esl_p_XY}.

\subsubsection{Fixed points}
We investigate the existence of fixed points in the averaged system \eqref{esl_p_XY}. Let $(X^*, Y^*)$ be a fixed point in Eq. \eqref{esl_p_XY}. 
Then, the coordinates are given by
\begin{subequations}
    \label{p_fix_XY}
    \begin{align}
        X^* &= \frac{2}{A} R \left( R^2 - R - \mu \right), \\
        Y^* &= \frac{2}{A} R \left[ b R + (a-\Omega) \right], 
    \end{align}
\end{subequations}
where $R$ is a non-negative root of the following 5th-order polynomial: 
\begin{multline}
    \label{p_f_def}
    f(R) \coloneqq R^5 - 2 R^4 + (1+b^2-2\mu)R^3 \\
    + 2[b(a-\Omega)+\mu]R^2 + [(a-\Omega)^2+\mu^2]R - \frac{A^2}{4}.
\end{multline}
See Appendix \ref{app_p_fix} for the derivation of Eqs. \eqref{p_fix_XY} and \eqref{p_f_def}.

\subsubsection{Phase plane and bifurcation analyses}
\label{sec_p_bif}
\begin{figure}
    \centering
    \includegraphics[width = 1.0\linewidth]{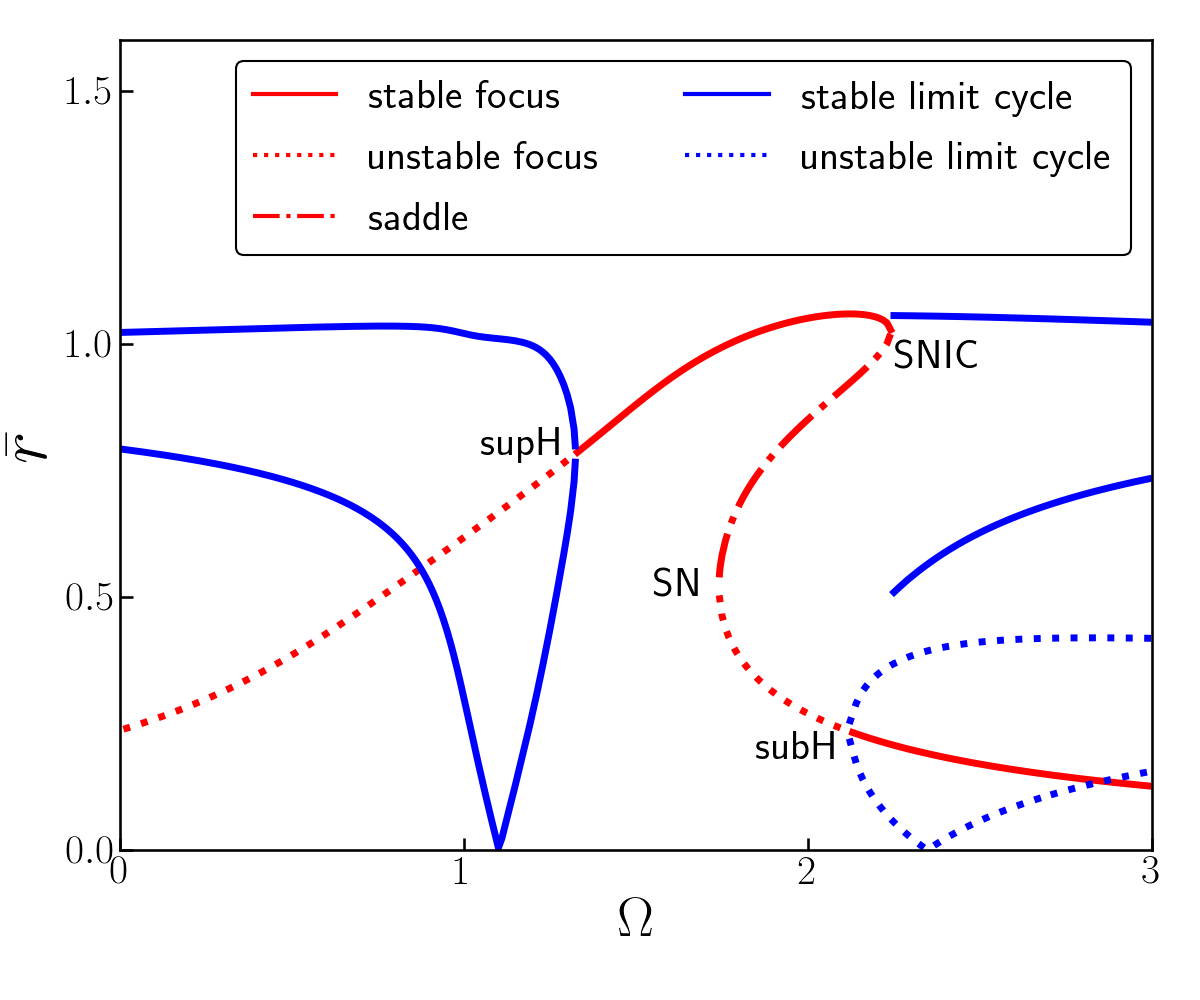}
    \caption{The bifurcation diagram of Eq. \eqref{esl_p_XY} when $A = 0.5$. We plot the amplitude $\bar{r}$ of equilibrium states against different values of $\Omega$. The red solid, dotted, and dash-dotted lines represent the stable focus, unstable focus, and saddle points, respectively. The blue solid lines show the maxima and minima of the amplitudes of stable limit cycles, whereas the blue dotted lines show those of unstable limit cycles. Combining with the results of phase plane analysis (Fig. \ref{esl_freq_pp_xy}), this figure suggests that 1:1 synchronization starts at $\Omega \simeq 1.11$  and ends at $\Omega \simeq 2.24$. SN: saddle-node bifurcation, supH: supercritical Hopf bifurcation, subH: subcritical Hopf bifurcation.}
    \label{esl_freq_bif}
\end{figure}

\begin{figure*}[ht]
    \centering
    \includegraphics[width = 1.\linewidth]{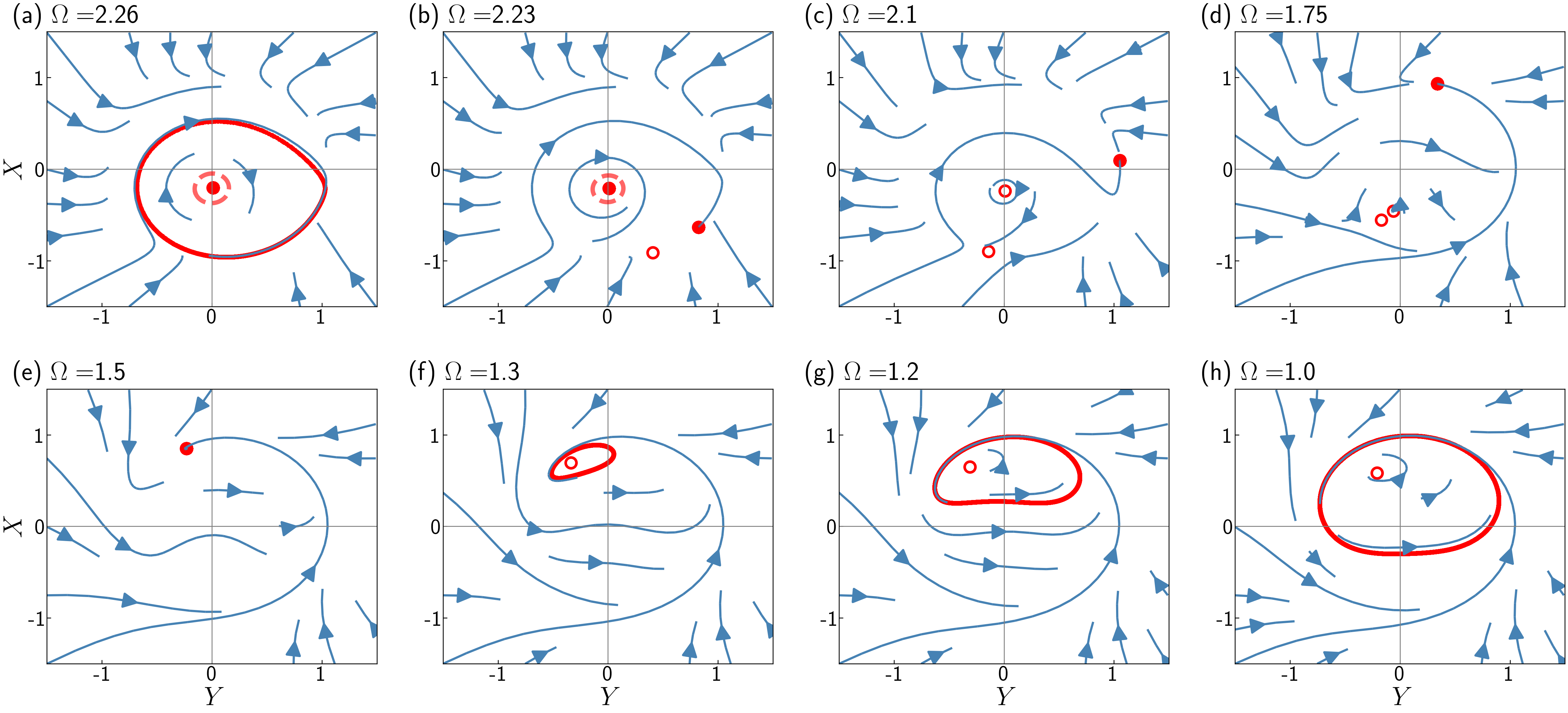}
    \caption{Phase planes of Eq. \eqref{esl_p_XY} for different values of $\Omega$. We fix $A=0.5$ in this figure. The blue lines show the flows. The red dots and red circles represent the stable and unstable fixed points, respectively. The red solid lines and red dashed lines denote the stable and unstable limit cycles, respectively. See main texts for the explanation of each panel [(a)-(h)]. }
    \label{esl_freq_pp_xy}
\end{figure*}

Here, we perform the phase plane and bifurcation analyses. Figures \ref{esl_freq_bif} and \ref{esl_freq_pp_xy} show the bifurcation diagram and typical phase planes of Eq. \eqref{esl_p_XY} for different values of $\Omega$, respectively. When investigating the stability of fixed points, we numerically solve the equation $f(R) = 0$ and calculate the eigenvalues of the corresponding Jacobian matrices. The stable and unstable limit cycles are derived by the numerical integration of Eq. \eqref{esl_p_XY} and its time-reversed system (i.e., Eq. \eqref{esl_p_XY} with a conversion $t \to -t$), respectively.  

These figures suggest that, for a sufficiently large $\Omega$, the averaged system \eqref{esl_p_XY} has a stable limit cycle, an unstable limit cycle, and a fixed point [Fig. \ref{esl_freq_pp_xy} (a)]. When $\Omega$ decreases, the stable limit cycle disappears by a SNIC (saddle-node on invariant circle) bifurcation and two fixed points emerge [Fig. \ref{esl_freq_pp_xy} (b)]. Then, the unstable limit cycle disappears by a subcritical Hopf bifurcation [Fig. \ref{esl_freq_pp_xy} (c)], after which the pair of two fixed points, one is a saddle and the other an unstable focus, vanishes by a saddle-node bifurcation [Figs. \ref{esl_freq_pp_xy} (d) and (e)]. Finally, a stable limit cycle appears by a supercritical bifurcation [Figs. \ref{esl_freq_pp_xy} (f)] and the limit cycle approaches a rotationally symmetric shape [Figs. \ref{esl_freq_pp_xy} (g) and (h)].

Remind that Eq. \eqref{esl_p_XY} is described on a rotation coordinate system. Namely, according to Eq. \eqref{comp_cv}, we see that the point $(X,Y)$ in system \eqref{esl_p_XY} rotates with angular velocity $\Omega$ in the $xy$-plane of the original system \eqref{esl_p}. Therefore, we find that the SNIC bifurcation between Fig. \ref{esl_freq_pp_xy} (a) and (b) corresponds to the onset of 1:1 synchronization with the periodic force. In other words, as shown in Figs. \ref{esl_freq_pp_xy} (a) and (b), the system state transits from a stable limit cycle (asynchronous state) to a stable fixed point (synchronous state). Even after the stable fixed point is destabilized by a supercritical Hopf bifurcation [Fig. \ref{esl_freq_pp_xy} (f)], 1:1 synchronization continues. This is because, as long as the system stays within a localized area in $XY$-plane, the average frequency measured in $xy$-plane matches $\Omega$. However, after the limit cycle crosses the origin, 1:1 synchronization no longer exists [Figs. \ref{esl_freq_pp_xy} (h)]. 

The above findings qualitatively agree with the direct simulation results of Eq. \eqref{esl_p}. Remind that, according to Fig. \ref{reso_esl_freq} (b), 1:1 synchronization occurs when $\Omega \in [\Omega_{\mathrm l}, \Omega_{\mathrm u}]$ with $\Omega_{\mathrm l} \simeq 1.12$ and $\Omega_{\mathrm u} \simeq 2.23$. On the other hand, Fig. \ref{esl_freq_bif} suggests that 1:1 synchronization starts at $\Omega \simeq 1.11$ (when the stable limit cycle crosses the origin) and ends at $\Omega \simeq 2.24$ (when the SNIC bifurcation occurs). 

In the following sections, we investigate the dynamics of a bistable oscillator under other types of periodic perturbations that diminish at the origin. Namely, we use multiplicative periodic forces combined with feedback terms. As we will discuss in the next two sections, the dynamics of oscillators are qualitatively different from those analyzed in this section. 

\section{Multiplicative periodic force: Case 1}
\label{sec_esl_mp1}
Here, we use a perturbation which is expressed as a product of a periodic force and first-order feedback. Namely, we consider
\begin{subequations}
    \label{esl_mp}
    \begin{align}
        \dot x &= \mu x - a y + (x^2 + y^2)(x - b y) - (x^2 + y^2)^2 x + A x \cos \Omega t, \label{esl_mp_x}\\
        \dot y &= a x + \mu y + (x^2 + y^2)(b x + y) - (x^2 + y^2)^2 y, 
    \end{align}
\end{subequations}
or 
\begin{equation}
    \label{esl_mp_comp}
    \dot z = (\mu + ia)z + (1 + ib)|z|^2 z - |z|^4 z + (\mathrm{Re}\: z)\: A \cos \Omega t.
\end{equation}

\subsection{Numerical simulation}
\begin{figure}[ht]
    \centering
    \includegraphics[width = 1.\linewidth]{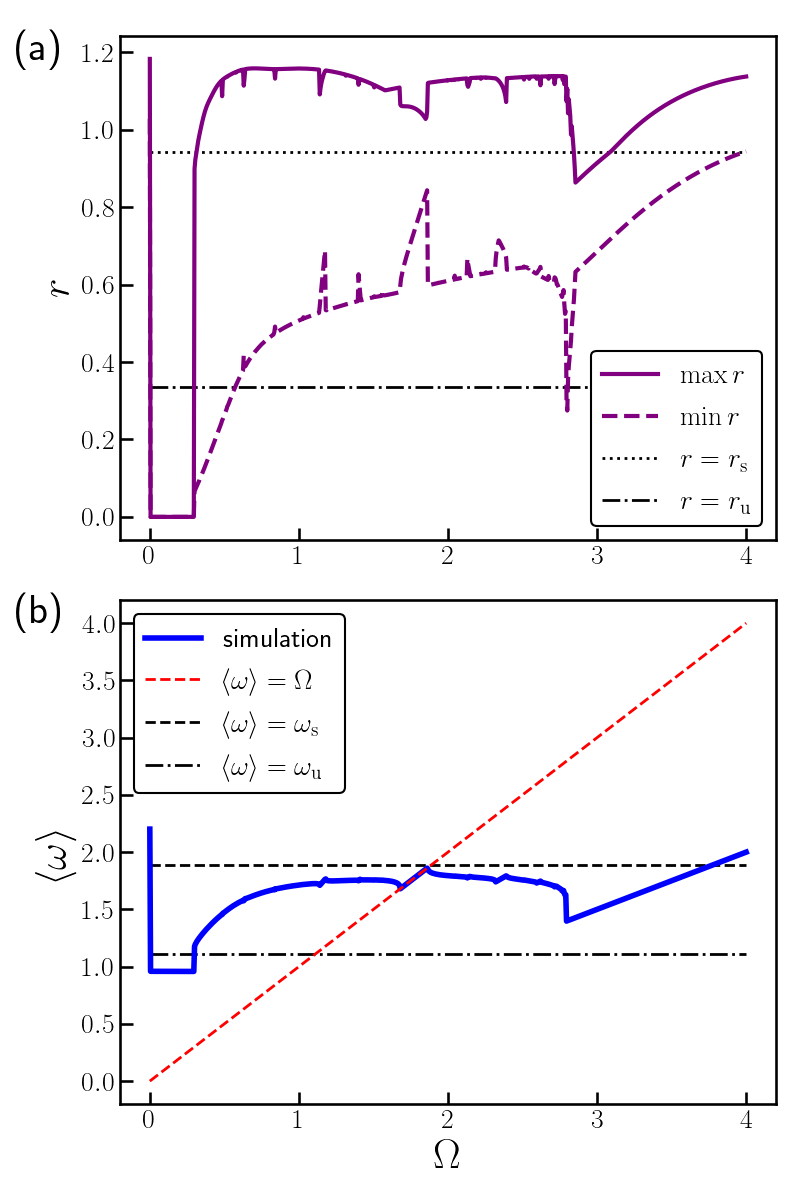}
    \caption{The amplitude (a) and frequency (b) response curves of Eq. \eqref{esl_mp}. The vertical and horizontal axes are the same as in Fig. \ref{reso_esl_freq}. Note that panel (b) suggests that the entrainment to the periodic force arises when $\Omega \in [\Omega_{\mathrm l}, \Omega_{\mathrm u}]$ with $\Omega_{\mathrm l} \simeq 1.68$ and $\Omega_{\mathrm u} \simeq 1.86$. Also note that, in panel (b), the frequency equals 1 if oscillation quenching occurs. This is because the frequency of the spiral at the origin is 1 [see Eq. \eqref{esl_theta}]. For numerical simulation, we set $A = 0.8, \: x(0)=3.0,\: y(0)=0.0$ and $t_{\rm stim} = 500$.}
    \label{reso_esl_multi_freq}
\end{figure}

Figure \ref{reso_esl_multi_freq} shows the amplitude and frequency response curves of Eq. \eqref{esl_mp}.

Figure \ref{reso_esl_multi_freq} (a) suggests that quenching occurs when $\Omega$ is small. We speculate the mechanism of this quenching as follows: if $\Omega$ is sufficiently small, then we can assume $\cos \Omega t \simeq -1$ during a sufficiently long time interval. In addition to this assumption, if we assume that $A$ is sufficiently large, then Eq. \eqref{esl_mp_x} approximately follows
\begin{equation}
    \dot x \simeq -A x. \label{esl_mp_approx}
\end{equation}
Since $x=0$ is a globally stable fixed point in Eq. \eqref{esl_mp_approx}, we see that quenching occurs if $\Omega$ is sufficiently small and $A$ is sufficiently large. 

Figure \ref{reso_esl_multi_freq} (b) suggests the existence of 1:1 synchronization. However, the interval of $\Omega$ for synchronization is much narrower than the previous case [Eq. \eqref{esl_p}]. We are going to analyze why 1:1 synchronization is unlikely to occur in system \eqref{esl_mp}. 

\subsection{Averaging}
To focus on the 1:1 synchronized state, we introduce a variable $Z$ given by Eq. \eqref{comp_cv}. Then, Eq. \eqref{esl_mp_comp} is rewritten as 
\begin{align}
    \dot{Z} & = [\mu + i(a-\Omega)] Z + (1 + ib)|Z|^2 Z - |Z|^4 Z \notag \\
    & \qquad + (\mathrm{Re}\: Z e^{i\Omega t}) e^{- i \Omega t} A \cos \Omega t \notag \\
    & = [\mu + i(a-\Omega)] Z + (1 + ib)|Z|^2 Z - |Z|^4 Z + A \cos \Omega t \notag \\
    & \qquad \times (\cos \Omega t \mathrm{Re}\: Z - \sin \Omega t \mathrm{Im}\: Z) (\cos \Omega t - i \sin \Omega t) \notag \\
    & = [\mu + i(a-\Omega)] Z + (1 + ib)|Z|^2 Z - |Z|^4 Z \notag \\
    & + \frac{A}{4} \left[ \left(3 \mathrm{Re}\: Z + i\: \mathrm{Im}\: Z\right) \cos \Omega t - \left(\mathrm{Im}\: Z + i\: \mathrm{Re}\: Z \right) \sin \Omega t \right. \notag \\
    & + \left. \left(\mathrm{Re}\: Z - i\: \mathrm{Im}\: Z \right) \cos 3 \Omega t - \left(\mathrm{Im}\: Z + i\: \mathrm{Re}\: Z \right) \sin 3 \Omega t \right] . \label{esl_mp_comp_cv}
\end{align}
We use $\mathrm{Re}\: \alpha \beta = \mathrm{Re}\: \alpha\: \mathrm{Re}\: \beta - \mathrm{Im}\: \alpha\: \mathrm{Im}\: \beta$ for the derivation of Eq. \eqref{esl_mp_comp_cv}. 

From Eq. \eqref{esl_mp_comp_cv}, we assume that the time series of $Z$ has a fast oscillation component of frequency $\Omega$. By averaging Eq. \eqref{esl_mp_comp_cv} over a period of the fast oscillation (i.e., $t \in [0, 2\pi/\Omega]$), we have
\begin{equation}
    \dot{Z} = [\mu + i(a-\Omega)] Z + (1 + ib)|Z|^2 Z - |Z|^4 Z. \label{esl_mp_comp_av}
\end{equation}
Note that Eq. \eqref{esl_mp_comp_av} is equivalent to the unperturbed system \eqref{esl}. Namely, by performing a lowest-order averaging, the perturbation term $A x \cos \Omega t$ vanishes. 

Equation \eqref{esl_mp_comp_av} suggests that 1:1 synchronization occurs only at $\Omega = \omega_{\rm s}$, which partly agrees with the simulation of original system \eqref{esl_p} in that 1:1 synchronization is observed in a small area near $\Omega = \omega_{\rm s} \simeq 1.88$ [Fig. \ref{reso_esl_multi_freq} (b)]. 
However, this analysis does not explain why 1:1 synchronization occurs in an interval of $\Omega$ (i.e., $[\Omega_{\mathrm l}, \Omega_{\mathrm u}]$ with $\Omega_{\mathrm l} \simeq 1.68$, $\Omega_{\mathrm u} \simeq 1.86$), not exactly at $\Omega = \omega_{\rm s}$. We expect that this dissociation results from the higher-order terms that we neglect in the averaging. In this paper, we do not further analyze this disagreement because this problem is not directly related to our study purpose, i.e., the investigation of state-transition methods. However, we consider that performing an averaging method considering higher-order terms is an important future challenge to further analyze the dynamics in Eq. \eqref{esl_p}. 

\section{Multiplicative periodic force: Case 2}
\label{sec_esl_mp2}
Next, we investigate the following system where a periodic force is combined with quadratic feedback $x^2$: 
\begin{subequations}
    \label{esl_mp2}
    \begin{align}
        \dot x &= \mu x - a y + (x^2 + y^2)(x - b y) - (x^2 + y^2)^2 x + A x^2 \cos \Omega t, \\
        \dot y &= a x + \mu y + (x^2 + y^2)(b x + y) - (x^2 + y^2)^2 y, 
    \end{align}
\end{subequations}
or 
\begin{equation}
    \label{esl_mp2_comp}
    \dot z = (\mu + ia)z + (1 + ib)|z|^2 z - |z|^4 z + (\mathrm{Re}\: z)^2\: A \cos \Omega t.
\end{equation}

\subsection{Numerical simulation}
\begin{figure*}
    \centering
    \includegraphics[width = 1.\linewidth]{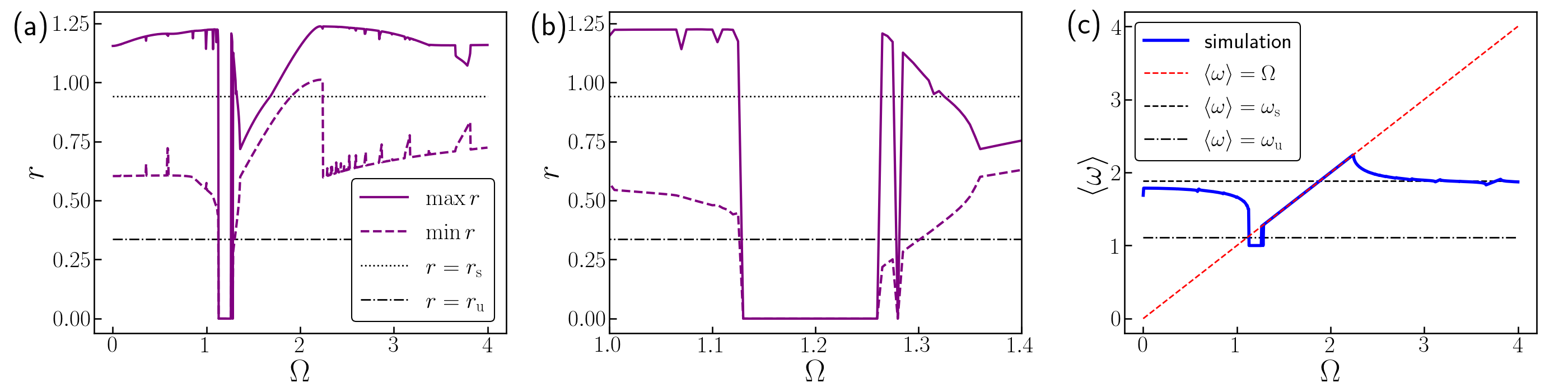}
    \caption{(a), (b) The amplitude response curves of Eq. \eqref{esl_mp2}. The maxima and minima of the amplitude $r$ are plotted against different values of $\Omega$. Panel (b) magnifies panel (a) to show the onset and offset of the region where oscillation quenching occurs. (c) The frequency $\langle \omega \rangle$ of Eq. \eqref{esl_mp2} is plotted against different values of $\Omega$. Panel (c) suggests that the entrainment to the periodic force occurs when $\Omega \in [\Omega_{\mathrm l}, \Omega_{\mathrm u}]$ with $\Omega_{\mathrm l} \simeq 1.285$ and $\Omega_{\mathrm u} \simeq 2.24$. For numerical simulation, we set $A = 1.0, \: x(0)=3.0,\: y(0)=0.0$ and $t_{\rm stim} = 500$.}
    \label{reso_esl_multi2}
\end{figure*}

Figure \ref{reso_esl_multi2} shows the amplitude and frequency response curves of Eq. \eqref{esl_mp2}. 

The amplitude response curves [Figs. \ref{reso_esl_multi2} (a) and (b)] suggest that oscillation quenching occurs at around $\Omega \simeq 1.20$. According to the frequency response curve [Fig. \ref{reso_esl_multi2} (c)], 1:1 synchronization occurs for some values of $\Omega$, including the region of $\Omega$ where oscillation quenching occurs. Based on these results, we expect that quenching is induced by the entrainment to the periodic force. 

\subsection{Analysis}
\subsubsection{Averaging}
We aim to analyze why oscillation quenching occurs, focusing on the 1:1 synchronized state between the oscillator and the periodic force. By introducing a variable $Z$ given by Eq. \eqref{comp_cv}, we rewrite Eq. \eqref{esl_mp2_comp} as 
\begin{align}
    \dot{Z} &= [\mu + i(a-\Omega)] Z + (1 + ib)|Z|^2 Z - |Z|^4 Z \notag \\
    & \qquad + (\mathrm{Re}\: Z e^{i\Omega t})^2\: e^{- i \Omega t} A \cos \Omega t, \notag \\
    &= [\mu + i(a-\Omega)] Z + (1 + ib)|Z|^2 Z - |Z|^4 Z + A \cos \Omega t \notag \\
    & \qquad \times (\cos \Omega t \mathrm{Re}\: Z - \sin \Omega t \mathrm{Im}\: Z)^2\: (\cos \Omega t - i \sin \Omega t). \notag \\
    &= [\mu + i(a-\Omega)] Z + (1 + ib)|Z|^2 Z - |Z|^4 Z \notag \\
    & + \frac{A}{8} \left\{ 3(\mathrm{Re}\: Z)^2 + (\mathrm{Im}\: Z)^2 + 2i\: \mathrm{Re}\: Z\: \mathrm{Im}\: Z + 4 (\mathrm{Re}\: Z)^2 \cos 2\Omega t \right. \notag \\
    & - 2 \left[2\mathrm{Re}\: Z\: \mathrm{Im}\: Z + i\: (\mathrm{Re}\: Z)^2 + i\: (\mathrm{Im}\: Z)^2 \right] \sin 2\Omega t \notag \\
    & \quad + \left[(\mathrm{Re}\: Z)^2 - (\mathrm{Im}\: Z)^2 - 2i\: \mathrm{Re}\: Z\: \mathrm{Im}\: Z \right] \cos 4\Omega t \notag \\
    & \qquad \left. - \left[2\mathrm{Re}\: Z\: \mathrm{Im}\: Z + i\: (\mathrm{Re}\: Z)^2 - i\: (\mathrm{Im}\: Z)^2 \right] \sin 4\Omega t \right\}.
    \label{esl_mp2_comp_cv}
\end{align}

From Eq. \eqref{esl_mp2_comp_cv}, we assume that the time series of $Z$ has a fast oscillation component of frequency $2\Omega$. By averaging Eq. \eqref{esl_mp2_comp_cv} over a period of the fast oscillation (i.e., $t \in [0, \pi/\Omega]$), we obtain
\begin{multline}
    \dot{Z} = [\mu + i(a-\Omega)] Z + (1 + ib)|Z|^2 Z - |Z|^4 Z \\
    + \frac{A}{8}\left[ 3(\mathrm{Re}\: Z)^2 + (\mathrm{Im}\: Z)^2 + 2i\: \mathrm{Re}\: Z\: \mathrm{Im}\: Z \right], \label{esl_mp2_comp_av}
\end{multline}
which is equivalent to 
\begin{subequations}
    \label{esl_mp2_XY}
    \begin{align}
        \dot X &= \mu X - (a-\Omega) Y + (X^2 + Y^2)(X - b Y) - (X^2 + Y^2)^2 X \notag \\
        & \qquad + \frac{A}{8} (3X^2 + Y^2), \\
        \dot Y &= (a-\Omega) X + \mu Y + (X^2 + Y^2)(b X + Y) - (X^2 + Y^2)^2 Y \notag \\
        & \qquad + \frac{A}{4} XY. 
    \end{align}
\end{subequations}

\subsubsection{Fixed points}
Let $(X^*, Y^*)$ be a fixed point in \eqref{esl_mp2_XY}. Then, the coordinates are given by either $(X^*, Y^*)=(0,0)$ or 
\begin{subequations}
    \label{mp2_fix_XY}
    \begin{align}
        X^* &= \frac{8}{3A} \left( R^2 - R - \mu \right), \\
        Y^* &= \frac{8}{A} \left[ b R + (a-\Omega) \right], 
    \end{align}
\end{subequations}
where $R$ is a non-negative root of the following 4th-order polynomial:
\begin{multline}
    \label{mp2_g_def}
    g(R) \coloneqq R^4 - 2 R^3 + (1+9b^2-2\mu)R^2 \\
    + [18b(a-\Omega)+2\mu-\frac{9A^2}{64}]R + [9(a-\Omega)^2+\mu^2].
\end{multline}
See Appendix \ref{app_mp2_fix} for the derivation of Eqs. \eqref{mp2_fix_XY} and \eqref{mp2_g_def}. 

\subsubsection{Phase plane and bifurcation analyses}
\begin{figure}[ht]
    \centering
    \includegraphics[width = 1.0\linewidth]{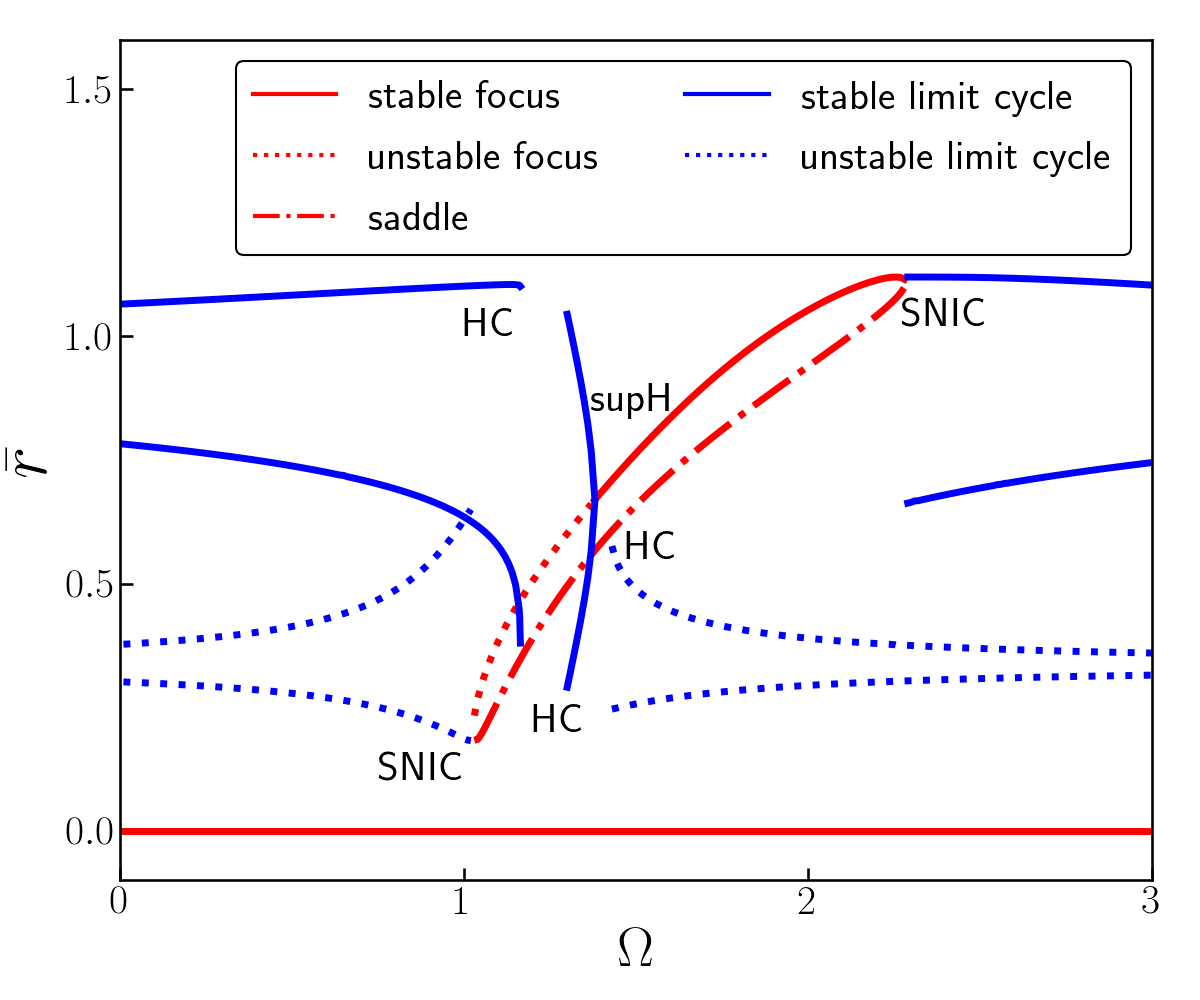}
    \caption{The bifurcation diagram of Eq. \eqref{esl_mp2_XY} when $A = 1.0$. We plot the amplitude $\bar{r} = |X+iY|$ of equilibrium states in Eq. \eqref{esl_mp2_XY} against different values of $\Omega$. The red solid, dotted, and dash-dotted lines represent the stable focus, unstable focus, and saddle points, respectively. The blue solid lines show the maxima and minima of the amplitudes of stable limit cycles, whereas the blue dotted lines show those of unstable limit cycles. Combining with the results of phase plane analysis (Fig. \ref{esl_multi2_pp_xy}), this figure suggests that 1:1 synchronization is observed between $\Omega \simeq 1.3$  and $\Omega \simeq 2.29$. HC: homoclinic bifurcation, supH: supercritical Hopf bifurcation.}
    \label{esl_freq_multi2_bif}
\end{figure}

\begin{figure*}[ht]
    \centering
    \includegraphics[width = 1.\linewidth]{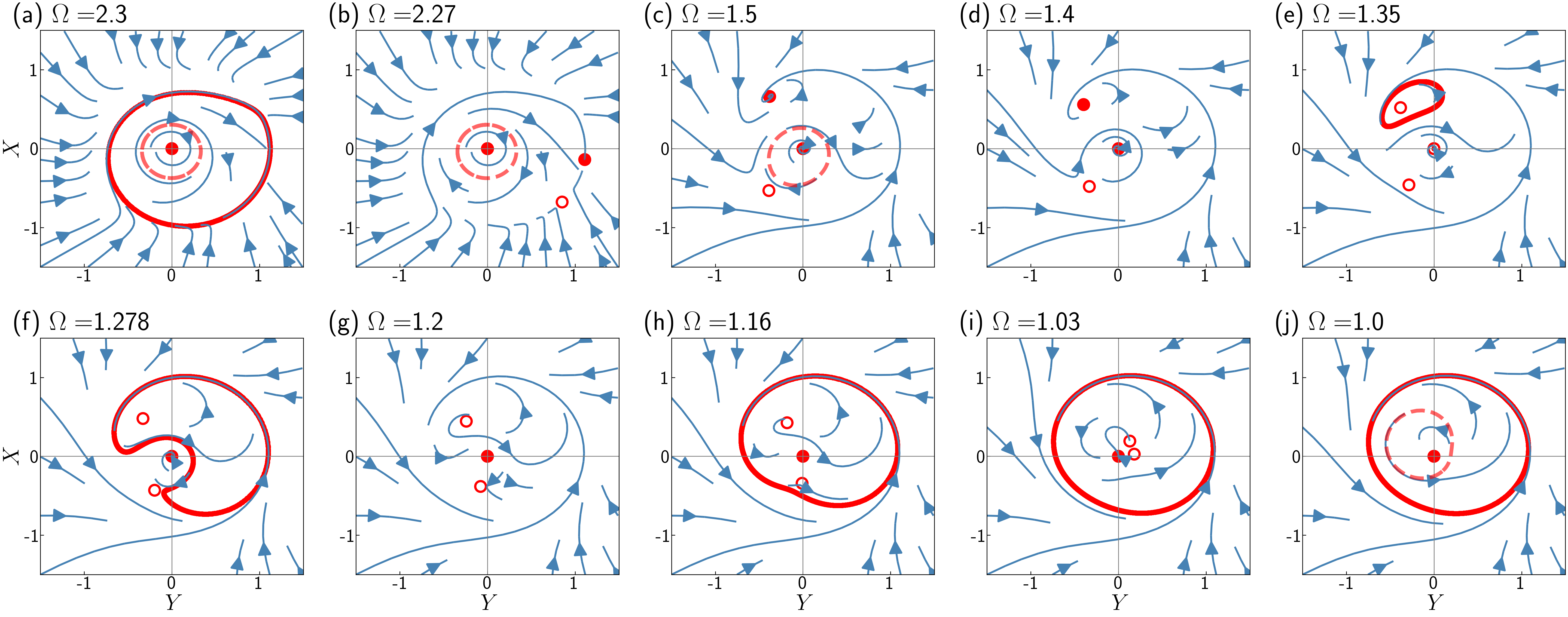}
    \caption{Phase planes of Eq. \eqref{esl_mp2_XY} for different values of $\Omega$. We fix $A=1.0$ in this figure. The blue lines show the flows. The red dots and red circles represent the stable and unstable fixed points, respectively. The red solid lines and red dashed lines denote the stable and unstable limit cycles, respectively. See main texts for the explanation of each panel [(a)-(h)]. }
    \label{esl_multi2_pp_xy}
\end{figure*}

Figures \ref{esl_freq_multi2_bif} and \ref{esl_multi2_pp_xy} show the bifurcation diagram and typical phase planes of Eq. \eqref{esl_mp2_XY} for different values of $\Omega$, respectively. As with the simulations in Sec. \ref{sec_p_bif}, we investigate the stability of fixed points by numerically solving the equation $g(R) = 0$ and calculating the eigenvalues of the corresponding Jacobian matrices. The stable and unstable limit cycles are obtained by the numerical integration of Eq. \eqref{esl_mp2_XY} and its time-reversed system (i.e., Eq. \eqref{esl_mp2_XY} with a conversion $t \to -t$), respectively.  


These figures suggest that, for a sufficiently large $\Omega$, the averaged system \eqref{esl_mp2_XY} has a stable limit cycle, an unstable limit cycle, and a fixed point [Fig. \ref{esl_multi2_pp_xy} (a)]. When $\Omega$ decreases, the stable limit cycle disappears by a SNIC bifurcation and two fixed points emerge [Fig. \ref{esl_multi2_pp_xy} (b)]. Then, the unstable limit cycle disappears by a homoclinic bifurcation [Figs. \ref{esl_multi2_pp_xy} (c) and (d)] and the stable fixed point destabilizes by a supercritical Hopf bifurcation [Fig. \ref{esl_multi2_pp_xy} (e)]. The stable limit cycle deforms its shape [Fig. \ref{esl_multi2_pp_xy} (f)] and vanishes by a homoclinic bifurcation [Fig. \ref{esl_multi2_pp_xy} (g)]. After that, a stable limit cycle emerges by a homoclinic bifurcation [Fig. \ref{esl_multi2_pp_xy} (h)]. Finally, the pair of two fixed points, one is a saddle and the other an unstable focus, vanishes by a SNIC bifurcation, accompanying an unstable limit cycle [Figs. \ref{esl_multi2_pp_xy} (i) and (j)]. 

Note that Eq. \eqref{esl_mp2_XY} is on a rotation coordinate system, seeing from the original system \eqref{esl_mp2}. We expect that 1:1 synchronization occurs between the SNIC bifurcation [between Figs. \ref{esl_multi2_pp_xy} (a) and (b)] and the homoclinic bifurcation [between Figs. \ref{esl_multi2_pp_xy} (e) and (f)]. Therefore, we can estimate from Fig. \ref{esl_freq_multi2_bif} that 1:1 synchronization occurs between $\Omega \simeq 1.3$ (the SNIC bifurcation point) and $\Omega \simeq 2.29$ (the homoclinic bifurcation point).
On the other hand, according to the direct simulation of the original system [Fig. \ref{reso_esl_multi2} (c)], 1:1 synchronization occurs when $\Omega \in [\Omega_{\mathrm l}, \Omega_{\mathrm u}]$ with $\Omega_{\mathrm l} \simeq 1.335$ and $\Omega_{\mathrm u} \simeq 2.185$. This interval is in good agreement with that estimated from Fig. \ref{esl_freq_multi2_bif}, which validates our averaging analysis.

It is noteworthy that the origin is the only stable attractor in Fig. \ref{esl_multi2_pp_xy} (f). This monostablity explains why oscillation quenching occurs in Fig. \ref{reso_esl_multi2} (b) around $\Omega \simeq 1.20$. The existence of a monostable attractor with a small amplitude is unique to this system \eqref{esl_mp2}, not observed in either Eq. \eqref{esl_p} or \eqref{esl_mp}. Thus, we consider that the perturbation used in Eq. \eqref{esl_mp2}, a periodic force combined with quadratic feedback, is useful for the annihilation of oscillation. 

\section{Discussion}
\label{sec_dis}
Exploring state-transition methods in a bistable oscillator is an important research topic that could lead to annihilating abnormal oscillations in the real world. 
Although periodically driven monostable oscillators have been addressed in previous studies, the amplitude response and state transition in a periodically driven bistable oscillator have yet to be investigated. 
In the present study, we analyze an extended Stuart-Landau oscillator driven by additive and multiplicative periodic forces. By focusing on the 1:1 frequency synchronization and performing the averaging approximation, we convert the non-autonomous model systems into autonomous systems and numerically analyze the bifurcation structures. We find that oscillation quenching occurs when we give a multiplicative periodic force combined with quadratic feedback. 

Among the three types of perturbations, oscillation quenching is observed when we add multiplicative periodic forces [Eqs. \eqref{esl_mp}, \eqref{esl_mp2}]. This result is understandable, considering that the origin [$(x,y)=(0,0)$] remains a stable fixed point in systems \eqref{esl_mp} and \eqref{esl_mp2}, but it is no more a fixed point in system Eq. \eqref{esl_p}. 
Our findings also indicate that the periodic force combined with nonlinear feedback [i.e., Eq. \eqref{esl_mp2}] induces oscillation quenching for some values of $\Omega$ that are not close to $0$. This is because, we consider, this perturbation in Eq. \eqref{esl_mp2} have constant terms that do not possess high-frequency components during the 1:1 synchronized state [compare Eq. \eqref{esl_mp2_comp_cv} with \eqref{esl_mp_comp_cv}] and thus does not average out over a period of the fast oscillation.

Various bifurcations in periodically driven oscillators have been revealed in previous studies \cite{mettin1993bifurcation,koch2024ghost}. These include the SNIC, Hopf, period-doubling, Neimark-Sacker bifurcations, and saddle-node bifurcation of cycles, etc. Several of these bifurcations are also observed in our bistable oscillator model (Figs. \ref{esl_freq_bif}, \ref{esl_freq_pp_xy}, \ref{esl_freq_multi2_bif}, and \ref{esl_multi2_pp_xy}). 
It is remarkable that, even after reducing the system by an averaging method, the autonomous systems \eqref{esl_p_XY} and \eqref{esl_mp2_XY} still have diverse bifurcation structures. We expect that the original systems [Eqs. \eqref{esl_p}, \eqref{esl_mp}, and \eqref{esl_mp2}] would possess much more complex and diverse bifurcations. A rigorous bifurcation analysis of these nou-autonomous systems is an important research theme for future study. 

In this study, we add perturbations to only one variable ($x$) among two system variables ($x,y$). Nevertheless, quenching is achieved by giving a multiplicative periodic force [Eq. \eqref{esl_mp2}]. This perturbation method can be applied to situations where we can observe and have access to only one variable among the whole system variables. An example of such situations is encountered in the FitzHugh-Nagumo model \cite{fitzhugh1961impulses,nagumo1962active}, in which one variable corresponds to the membrane potential that we can monitor and perturb while the other is a hidden variable that is difficult to observe. We believe that our method, annihilating oscillations by intervening to only a part of system variables, enhances the potential for practical applications. 

By using an additive periodic force [Eq. \eqref{esl_p}], we find that the amplitude of oscillation can decrease below $r_{\mathrm u}$, the radius of the unstable limit cycle [Fig. \ref{reso_esl_freq} (a)]. This means that, by choosing the appropriate frequency of external force, the system state temporarily falls within the basin of the fixed point of the original system \eqref{esl}. Therefore, we expect that the following additive periodic force combined with feedback can achieve oscillation quenching: 
\begin{subequations}
    \label{esl_p_switch}
    \begin{align}
        \dot x &= \mu x - a y + (x^2 + y^2)(x - b y) - (x^2 + y^2)^2 x \notag \\
        & \qquad + A H(r(t) - r_{\rm u}) \cos \Omega t,\\
        \dot y &= a x + \mu y + (x^2 + y^2)(b x + y) - (x^2 + y^2)^2 y,
    \end{align}
\end{subequations}
where $H(\cdot)$ denotes the Heaviside step function. The perturbation term in Eq. \eqref{esl_p_switch} diminishes when the system falls within the basin of the fixed point (i.e., when $r(t) < r_{\rm u}$). Thus, we expect that oscillation quenching occurs in system Eq. \eqref{esl_p_switch}. This method can be applied to situations where we can observe a quantity that reflects the state of the entire system [$r(t)$] but have access to only a part of system variables [$x(t)$]. An example close to such a situation is an Implantable Cardioverter Defibrillator (ICD), which is a medical device implanted in the heart of patients with lethal cardiac arrhythmias (e.g., ventricular fibrillation). A lead of an ICD can monitor the activity of the heart and give an electrical pulse if necessary. An ICD often performs the monitoring and pulsing in different parts of the heart \cite{dimarco2003implantable} and thus is similar to the situation 
where the perturbation method in Eq. \eqref{esl_p_switch} might be helpful. 
Compared with the multiplicative periodic force [Eq. \eqref{esl_mp2}], the additive force [Eq. \eqref{esl_p_switch}] has an advantage in that we just have to monitor the system state, not using the observed values to change the strength of perturbation. We expect that an additive periodic force combined with feedback also has the potential for real-world applications. 

Several open questions remain to be examined. 
First, Fig. \ref{reso_esl_all} in Appendix \ref{app_sync} suggests the existence of $m:n$ synchronization, which we do not focus on in this study. Expanding our averaging analysis for the $m:n$ synchronization is an important task that helps further clarify the dynamics of periodically driven bistable oscillators. 
Second, when considering a periodic force combined with linear feedback [Eq. \eqref{esl_mp}], the perturbation averages out by the lowest-order averaging method. This is inconsistent with numerical results, in which the amplitude of the oscillator varies during 1:1 synchronization [Fig. \ref{reso_esl_multi_freq} (a)]. This disagreement between analysis and numerical simulation would be solved by an averaging method considering the effects of higher-order terms. 
Third, as we have mentioned above, the original nou-autonomous systems \eqref{esl_p}, \eqref{esl_mp}, and \eqref{esl_mp2} are expected to have complex and diverse dynamics, including quasi-periodic solutions and chaos. Bifurcation analyses of these systems would be challenging. 

In conclusion, we analyze the amplitude response of a periodically driven extended Stuart-Landau oscillator. We perform an averaging approximation and investigate the bifurcation structure. We find that a periodic force combined with quadratic feedback can induce oscillation quenching. We believe that this study, as well as our previous study that addressed feedback-induced quenching \cite{kato2024weakly}, is an important step for the systematic understanding of state transition methods in a bistable oscillator, and thus could have a potential for practical applications in terminating undesirable oscillations. 

\begin{acknowledgments}
This study was supported by JSPS KAKENHI (No. JP23KJ0756) to Y.K and JSPS KAKENHI (No. JP21K12056, No. JP22K18384, and No. JP23H02796) to H.K.
\end{acknowledgments}

\section*{Conflict of Interest}
The authors have no conflicts to disclose.

\section*{Data Availability Statement}
The data that support the findings of this study are available within the article.

\appendix
\section{Derivation of fixed points in Eq. \eqref{esl_p_XY}}
\label{app_p_fix}
Let $Z^*$ be a fixed point in Eq. \eqref{esl_p_comp_av}. By introducing a polar coordinate $r^* e^{i \theta^*} \coloneqq Z^*$, one has
\begin{equation}
    [\mu + i(a-\Omega)] r^* e^{i \theta^*} + (1 + ib)(r^*)^3 e^{i \theta^*} -(r^*)^5 e^{i \theta^*} = - \frac{A}{2}, 
    \label{esl_p_comp_fix}
\end{equation}
which implies that
\begin{subequations}
    \label{esl_p_polar_lock}
    \begin{align}
        (r^*)^5 - (r^*)^3 - \mu r^*
        &= \frac{A}{2}\cos \theta^*, \label{esl_p_polar_r_lock}\\
        b (r^*)^3 + (a - \Omega)r^*
        &= \frac{A}{2} \sin \theta^*. \label{esl_p_polar_theta_lock}
    \end{align}
\end{subequations}
By squaring both sides and summing Eqs. \eqref{esl_p_polar_r_lock} and \eqref{esl_p_polar_theta_lock}, we obtain
\begin{equation}
    f((r^*)^2) = 0, \label{reso_esl_avr_fix_R} 
\end{equation}
where $f$ is given by Eq. \eqref{p_f_def}. It thus follows from Eq. \eqref{esl_p_polar_lock} that 
\begin{align}
    X^* &= \mathrm{Re}\: Z^* = r^* \cos \theta^* = \frac{2}{A} (r^*)^2 [(r^*)^4 - (r^*)^2 - \mu], \\
    Y^* &= \mathrm{Im}\: Z^* = r^* \sin \theta^* = \frac{2}{A} (r^*)^2 [b (r^*)^2 + (a-\Omega)].
\end{align}
We set $R \coloneqq (r^*)^2$. Then, Eq. \eqref{p_fix_XY} holds. 

\section{Derivation of fixed points in Eq. \eqref{esl_mp2_XY}}
\label{app_mp2_fix}
Let $Z^*$ be a fixed point in Eq. \eqref{esl_mp2_comp_av}. We introduce a polar coordinate $r^* e^{i \theta^*} \coloneqq Z^*$. Noting that 
\begin{align*}
& 3(\mathrm{Re}\: Z)^2 + (\mathrm{Im}\: Z)^2 + 2i\: \mathrm{Re}\: Z\: \mathrm{Im}\: Z \\
= & (3 \mathrm{Re}\: Z - i\: \mathrm{Im}\: Z)(\mathrm{Re}\: Z + i\: \mathrm{Im}\: Z) = (3 \mathrm{Re}\: Z - i\: \mathrm{Im}\: Z)Z,
\end{align*}
one has
\begin{multline}
    [\mu + i(a-\Omega)] r^* e^{i \theta^*} + (1 + ib)(r^*)^3 e^{i \theta^*} -(r^*)^5 e^{i \theta^*} \\
    + \frac{A}{8}(3 r^* \cos \theta^* - i r^* \sin \theta^*) r^* e^{i \theta^*} = 0.
    \label{esl_mp2_comp_fix}
\end{multline}
Equation \eqref{esl_mp2_comp_fix} implies that
\begin{subequations}
    \label{esl_mp2_polar_lock}
    \begin{align}
        \frac{1}{3} [(r^*)^5 - (r^*)^3 - \mu r^*] &= \frac{A}{8} (r^*)^2 \cos \theta^*, \label{esl_mp2_polar_r_lock}\\
        b (r^*)^3 + (a - \Omega)r^* &= \frac{A}{8} (r^*)^2 \sin \theta^*. \label{esl_mp2_polar_theta_lock}
    \end{align}
\end{subequations}
By squaring both sides and summing Eqs. \eqref{esl_mp2_polar_r_lock} and \eqref{esl_mp2_polar_theta_lock}, we obtain
\begin{equation}
    (r^*)^2 g((r^*)^2) = 0, \label{esl_mp2_avr_fix_R} 
\end{equation}
where $g$ is given by Eq. \eqref{mp2_g_def}. Equation \eqref{esl_mp2_avr_fix_R} has a trivial solution $r^* = 0$ that corresponds to $(X^*, Y^*) = (0,0)$.  If $r^* \neq 0$, then Eq. \eqref{mp2_fix_XY} with $R \coloneqq (r^*)^2$ follows from Eqs. \eqref{esl_mp2_polar_lock} and the relationships $X^* = r^* \cos \theta^*$, $Y^* = r^* \sin \theta^*$.

\section{$m:n$ sunchronization}
\label{app_sync}
\begin{figure*}[h]
    \centering
    \includegraphics[width = 1.0\linewidth]{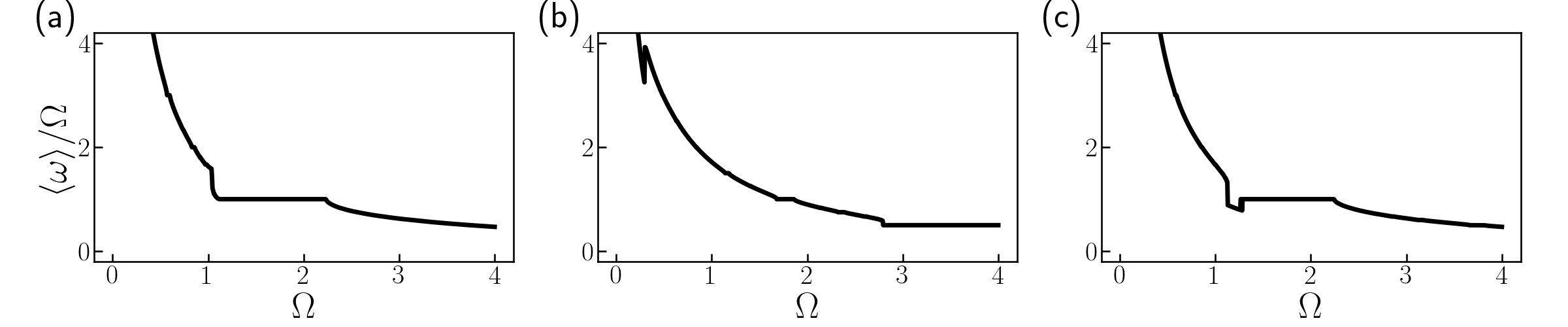}
    \caption{Panels (a), (b), and (c) represent the frequency ratio $\langle \omega \rangle/\Omega$ in system \eqref{esl_p}, \eqref{esl_mp}, and \eqref{esl_mp2}, respectively. Several $m:n$ synchronization states are observed. }
    \label{reso_esl_all}
\end{figure*}

In Fig. \ref{reso_esl_all}, we plot the frequency ratio of the oscillator to the periodic force (i.e., $\langle \omega \rangle/\Omega$) in system \eqref{esl_p} [panel (a)], \eqref{esl_mp} [panel (b)], and \eqref{esl_mp2} [panel (c)]. Note that $m:n$ synchronization states, other than 1:1 synchronization which we analyze in this study, are observed. 


\bibliography{papers}
\end{document}